\documentclass[12pt]{article}
\usepackage{graphicx}
\usepackage{float}
\def\reference{\parskip 0pt\par\noindent\hangindent 0.5 truecm}

\begin{document}
%
% Title
% Capitalise the title normally - do not use ALL CAPS.
%
\title{Radio Sources in the 2dF Galaxy Redshift Survey. 
I. Radio source populations\footnote {Based
on data obtained by the 2dFGRS Team: J Bland--Hawthorn (AAO), R D Cannon 
(AAO), S Cole (Durham), M Colless (ANU, Australian convenor), C A Collins
(Liverpool J Moores), W J Couch 
(UNSW), N Cross (St Andrews), G B Dalton (Oxford), K E Deeley
(UNSW/AAO), R De Propris (UNSW), S P Driver (St Andrews), G Efstathiou
(Cambridge), R S Ellis (Cambridge, UK convenor), S Folkes (Cambridge), 
C S Frenk (Durham), K Glazebrook (AAO),  N J Kaiser (Hawaii), O Lahav
(Cambridge), I J Lewis (AAO), S L Lumsden (Leeds), S J Maddox
(Cambridge), S Moody (Cambridge), P Norberg (Durham), J A Peacock
(Edinburgh), B A Peterson (ANU), I A Price (ANU), S Ronen (Cambridge),
M Seabourne (Oxford), R Smith (Edinburgh), W J Sutherland (Oxford), H
Tadros (Oxford), K Taylor (AAO).} }
%
% Authors
% Here comes the author(s) of the paper. Please add the appropriate author
% names for your paper and indicate within the $^...$ the number(s)
% which corresponds to the institute(s) of each author. In this example
% the second author has two institutional affiliations.
% Add or remove authors as required, maintaining the \and syntax between
% each author, but no \and after the last author.
% **** IMPORTANT: Leave the closing curly bracket line as is. ******

\author{Elaine M. Sadler $^{1}$ \and
 V.J. McIntyre $^{1}$ \and
 C.A. Jackson $^{1}$ \and
 R.D. Cannon $^{2}$
} % IMPORTANT: leave this curly bracket as the first character of this line.

% Date - leave this blank.
\date{}
\maketitle

% Institutions
% Here fill in your institute name(s) and address(es)
% The number in $^...$ indicates the author number.  For example
{\center
$^1$ School of Physics, University of Sydney, NSW 2006, Australia \\
ems@physics.usyd.edu.au, vjm@physics.usyd.edu.au, cjackson@physics.usyd.edu.au 
\\[3mm]
$^2$ Anglo--Australian Observatory, PO Box 296, Epping, NSW 2121, Australia \\ 
rdc@aaoepp.aao.gov.au \\[3mm]
}

% Abstract
% Simply place your abstract between the \begin{abstract} and
% \end{abstract} commands.
%
\begin{abstract}
We present the first results from a study of the radio continuum properties 
of galaxies in the 2dF Galaxy Redshift Survey, based on thirty 2dF fields 
covering a total area of about 100\,deg$^2$.  About 1.5\% of galaxies with
$b_{\rm J} < 19.4$ mag.\ are detected as radio continuum sources in the NRAO
VLA Sky Survey (NVSS).  Of these, roughly 40\% are star--forming galaxies and
60\% are active galaxies (mostly low--power radio galaxies and a few Seyferts).
The combination of 2dFGRS and NVSS will eventually yield a homogeneous 
set of around 4000 radio--galaxy spectra, which will be a powerful tool for
studying the distribution and evolution of both AGN and starburst galaxies out
to $z\sim0.3$. 
\end{abstract} 

{\bf Keywords:} 
% Place keywords here.  PASA uses the standard list of subject 
% headings adopted by The Astrophysical Journal and available from URL:
%   http://www.noao.edu/apj/keywords96.html

radio continuum: galaxies, galaxies: distances and redshifts, 
galaxies: active, galaxies: starburst

% A formatting command to add space between the author list and the body
% of the paper when printed. This spacing may be changed as desired.
\bigskip

%
% Body of paper
%

\section{Introduction}  
A new generation of sensitive, large--area radio--source surveys at 
milliJanksy levels (1\,Jy = 10$^{-26}$ W\,m$^{-2}$Hz$^{-1}$) is now 
becoming available.  They include the FIRST survey (Becker et al.\ 1995), 
WENSS (Rengelink et al.\ 1997), NVSS (Condon et al.\ 1998) and SUMSS (Bock 
et al.\ 1999).  These surveys offer some important advantages for cosmological
studies -- they reach sufficiently high source densities that detection of
large--scale structure is possible (Cress et al.\ 1996, Magliocchetti et 
al.\ 1998), and also probe a second cosmologically--significant radio source
population, that of star--forming galaxies, which are rarely seen in
strong--source surveys. 

Deep radio surveys of a few small areas of sky at 1.4\,GHz (Condon 1984, 
Windhorst et al.\ 1985; see also Condon 1992) have shown that there are two 
distinct populations of extragalactic radio sources.  
Over 95\% of radio sources above about 50\,mJy are classical radio galaxies 
and quasars (median redshift z$\sim$1) powered by active galactic nuclei  
(AGN), while the remainder are identified with star--forming galaxies (median
z$\sim$0.1).  The fraction of star--forming galaxies increases rapidly below
10\,mJy, and below 1\,mJy they begin to be the dominant population. 

The scientific return from radio continuum surveys is enormously increased 
if the optical counterparts of the radio sources can be identified and their
redshift distribution measured.  In the past, this has been a slow and tedious
process which could only be carried out for relatively small samples.  
However, the Anglo--Australian Observatory's Two--degree Field (2dF)
spectrograph now makes it possible to carry out spectroscopy of
several hundred galaxies simultaneously.  Here, we describe the first step in
this process -- the identification of faint radio--source counterparts among
galaxies whose spectra have been obtained in the 2dF Galaxy Redshift Survey
(2dFGRS). 

The 2dFGRS (Colless 1999, Maddox 1998) is a large--scale survey of 250,000
galaxies covering 2000 square degrees in the southern hemisphere.  The survey
is designed to be almost complete down to a limiting apparent magnitude of
$b_{\rm J}=19.4$. The median redshift of the galaxies is about 0.1 and the
great majority have $z < 0.3$.  Spectra are being obtained using the 2dF
multi--object fibre optic spectroscopic system on the 3.9m Anglo--Australian
Telescope (AAT) (Lewis et al.\ 1998; Smith \& Lankshear 1998).  2dF enables the
AAT to obtain spectra for 400 objects simultaneously, spread over a field which
is two degrees in diameter. The survey will cover two large strips of sky at
high Galactic latitude, one each in the southern and northern Galactic
hemispheres, plus outlying random fields in the south.  The first test data
were obtained in 1997 and the survey is expected to be substantially complete
by the end of 2000. 

This paper presents the first results from what will eventually be a much larger
study.  When the 2dFGRS is complete, it will yield around 4000 good--quality
spectra of galaxies associated with faint radio sources --- by far the largest
sample of radio--galaxy spectra ever compiled.  Our aim in this paper is to
present a first, qualitative exploration of the faint radio source population
as observed by 2dF and the NVSS. 

Throughout this paper, we use H$_o$=75 km/s/Mpc and q$_o$=0.5. 

\section{The optical data}  
Our data set comprises the thirty fields observed by the 2dFGRS team in 
November 1997 and January 1998.  Twenty--three of these are in the southern 
Galactic hemisphere and the other seven in the north. They include 
a total of 8362 target galaxies brighter than $b_{\rm J}$ magnitude 19.4.

Although each 2dF field covers about 3.14 square degrees, the total effective
area covered by the 30 fields is somewhat less than 90 square degrees because
there is some overlap between adjacent fields.  Also, the fractional
completeness of this early sample varies substantially from field to field.  
In many fields about 30\% of the fibres were allocated to targets in a 
parallel QSO survey; the 2dFGRS uses a flexible tiling algorithm to deal with
this and with the intrinsic variations in target density across the sky.  When
complete, the GRS will yield spectra for around 95\% of all galaxies in the
input catalogue.  The current sample has variable levels of completeness (in
terms of the surface distribution of all galaxies brighter than $b_{\rm J} =
19.4$ mag.), though there should be no systematic effects depending on the
magnitude, redshift or other properties of the galaxies. 

The standard observing pattern for the 2dFGRS is a set of three consecutive 
20--minute exposures per field, together with calibration arcs and flat fields.
The total exposure time is well--matched to the time required to reconfigure a
second set of fibres for the next observation.  Up to 380 galaxies can be 
observed simultaneously, with some 20 fibres allocated to sky.  In many fields,
however, the total number of galaxies is closer to 250 since a deep QSO survey
is being carried out in parallel with the galaxy redshift survey. 

Using 300 lines/mm gratings, the 2dFGRS spectra cover the wavelength range
3800\AA\ to 8000\AA\ at a resolution of about 10\AA.  Most spectra have a
signal--to--noise ratio (S/N) of 10 (per 4\AA\ pixel) or better.  Reliable
redshifts are obtained for up to 95\% of the targets in good observing
conditions; the survey average is currently about 90\% (Folkes et al.\ 1999). 
About 5\% of the targets are found to be foreground stars; the original
selection was for objects which appeared non--stellar in digitised data from 
UK Schmidt Telescope sky survey photographs, using a conservative criterion 
to minimise the number of galaxies missed. 

\begin{figure}[htb]
\centering
\includegraphics[width=0.7\textwidth]{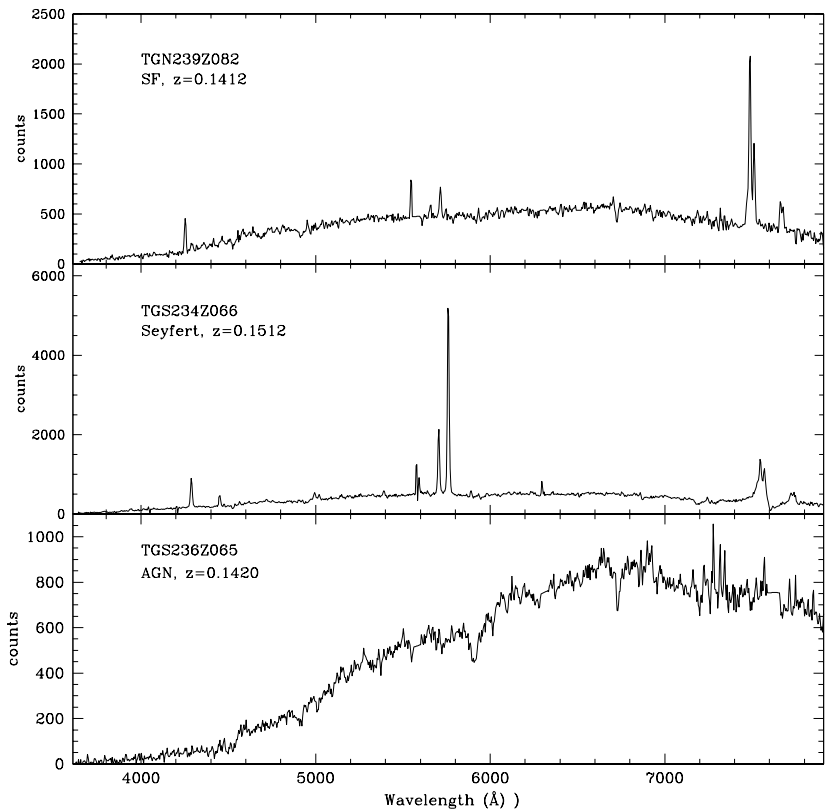}
\caption{Spectra of three 2dFGRS radio sources, showing the 
three main spectral classes: (top) A star--forming galaxy, TGN239Z082, 
with strong Balmer emission lines of H$\alpha$\ $\lambda6563$ and 
H$\beta$\ $\lambda4861$, as well as weaker emission lines of 
$[$SII$]$ $\lambda$6716,6731, $[$NII$]$ $\lambda$6583, $[$OIII$]$
$\lambda$4959,5007 and $[$OII$]$ $\lambda$3727; 
(middle) an emission--line AGN, TGS234Z066, with a Seyfert--like spectrum 
which has broad H$\alpha$ emission and $[$OIII$]$ much stronger than H$\beta$; 
(bottom) A radio--galaxy, TGS236Z065, with an absorption--line spectrum typical
of giant elliptical galaxies.}
\label{fig:spectra}
\end{figure}

Figure~\ref{fig:spectra} shows three spectra from the survey, and gives an idea of the 
typical quality of 2dFGRS spectra for galaxies with redshifts around $z=0.15$  
and $b_{\rm J}$ magnitudes in the range 17.6--18.9. 

\section{The radio data} 
The NRAO VLA Sky Survey (NVSS; Condon et al.\ 1998) is a 1.4\,GHz (20\,cm) 
radio imaging survey of the entire sky north of declination $-40^\circ$.  
The survey catalogue contains sources as weak as 2.5\,mJy, and is essentially
complete above 3.5\,mJy. 

We used the NVSS source catalogue to identify candidate radio--emitting 
galaxies in the 2dFGRS.  Subsets of the NVSS catalogue were extracted to 
match the RA and Dec range covered by each of the 2dF fields. 
The NVSS source density is roughly 60 per square degree, so each of
these sub-catalogues contained about 200 radio sources.  At this stage, we 
did not attempt to identify `double' NVSS sources.  However, we estimate 
(using an algorithm similar to that adopted by Magliocchetti et al.\ (1998) 
for identifying double radio sources in the FIRST survey) that these represent
only a small fraction of NVSS sources (of order $1$\%), so the presence of
double sources does not significantly affect our results in this small sample. 

We compared the NVSS and 2dFGRS catalogues for each field and identified the
galaxies for which there is a candidate radio `match', i.e. an NVSS radio
source lying within 15 arcsec of the position of a 2dF galaxy.  The 15 arcsec
limit was chosen because our earlier Monte Carlo tests using the COSMOS
database suggested that most candidate matches of {\it bright} galaxies (i.e.
$b_{\rm J} < $19.4 mag) with radio--optical separation up to 10 arcsec are real
associations, together with a substantial fraction of those with offsets of
10--15 arcsec. 

The uncertainty in the NVSS radio source positions increases from a 1$\sigma$
error of 1--2 arcsec at 10mJy to 4--5 arcsec (and occasionally up to 10 arcsec)
for the faintest (2--3mJy) sources (Condon et al.\ 1998).  In addition,
centroiding of bright radio sources with extended radio structure can be
somewhat uncertain and may make optical identification difficult.  
Determining the optical centroid also becomes imprecise for large nearby
galaxies. In these cases, overlaying the radio contours on an optical image
usually makes it clear whether or not the candidate ID is correct. 

We found a total of 127 candidate matches in the 30 fields studied, 
i.e. 1.5\% of the 8362 2dF galaxies in the survey area.  Of these, 99 had
radio--optical offsets of less than 10 arcsec. 

We also ran the matching program twice more, offsetting all the radio
positions by 3 and 5 arcmin.  Any matches produced from this
`off--source' catalogue should be chance coincidences, allowing us to
estimate the number of matches expected purely by chance.  Table~\ref{tab:offsets}
lists the results of this test, and shows the distribution of the
offset D (difference between radio and optical positions) for matches
found in the `on--source' and `off--source' tests.  

A further check
comes from predicting the expected number of chance coincidences,
based on the average surface densities of objects.  
Since 15 arcsec is 1/240 of the 1\,degree radius of each 2dF field, and
there are about 60 NVSS sources per square degree, the chance that a
given GRS target will fall within 15 arcsec of an unrelated radio
source is about 60$\times\pi$/240$^2$ (the resolution of the NVSS is such that
at most one source can be identified with each optical galaxy).  Thus
about 27 chance coincidences are expected in a total of 8362 galaxies 
(see column 5 of Table~\ref{tab:offsets}). 
This calculation ignores the known clustering of galaxies on the sky,
although this will only invalidate the result if there is significant
clustering on scales comparable to the identification range of 15
arcsec, or if there are substantial differences between the spatial
distribution of radio--loud and quiet galaxies.  We believe these
effects are unlikely to give an error of more than 10\%.

The results suggest that candidate matches with an offset of up to 
10\,arcsec are highly likely to be real associations.  We therefore use a
simple 10\,arcsec cutoff in radio--optical position difference for the analysis
which follows.  This gives us 99 radio--detected galaxy matches in the 30 2dF
fields. It also means that we have probably omitted about a dozen real
identifications with larger offsets, but this is not a problem here since our
aim is to make a first qualitative exploration of the faint radio galaxy
population. 

\begin{table}[H]
\begin{center}
\begin{tabular}{crrrrc} 
\hline
          D    &\multicolumn{3}{c}{No. of matches} & Predicted & Probability that \\
          (arcsec)     &      On  & Off\,3 &  Off\,5 & counts &  match is real \\
\hline
          0--2.5       &      34  &   3  &   0   &  1 & 96\% \\
        2.5--5.0       &      29  &   2  &   1   &  2 & 95\% \\
        5.0--10.0      &      36  &   6  &   8   &  9 & 81\% \\
       10.0--15.0      &      28  &  12  &  13   & 15 & 55\% \\
\hline
%&&&\\
\end{tabular}
\caption{Comparison of on--source and off--source matches as a function of 
radio--optical offset D.  `Off-source' data are for matches where the 
radio--source positions were offset in declination by 3\,arcmin (Off\,3) 
and 5\,arcmin (Off\,5).  }
\label{tab:offsets}
\end{center}
\end{table}

\section{Two kinds of radio source: AGN/SF classification} 
We classified each matched galaxy as either `AGN' or `star--forming'
(SF) based on its 2dF spectrum. `AGN' galaxies have either a pure
absorption--line spectrum like that of a giant elliptical galaxy, or a
stellar continuum plus nebular emission lines such as [OII] and [OIII]
which are strong compared with any Balmer--line emission.  Some of 
the emission--line AGN have spectra which resemble Seyfert galaxies.  
`SF' galaxies are those where strong, narrow emission lines of H$\alpha$ 
and (usually) H$\beta$ dominate the spectrum.  They include both nearby
spirals and more distant IRAS galaxies.  Figure~\ref{fig:spectra} shows examples of
spectra we classified as AGN, Seyfert and SF.  Note that in this
classification scheme, `AGN' may simply denote the presence of radio
emission, with no obvious optical signature.  The origin of radio emission 
in the AGN and SF galaxies is believed to be quite different (e.g. Condon 
1989), arising from non--thermal processes related to a central massive object
in the AGN galaxies and from processes related to star formation (including
supernova remnants, HII regions, etc.) in the SF galaxies. 

We are confident that this simple `eyeball' classification of the 2dF spectra 
allows us to separate the AGN and SF classes accurately.  
Jackson \& Londish (1999) measured several emission--line ratios 
(including [OIII, $\lambda$5007]/H$\beta$, [NII, $\lambda$6584]/H$\alpha$,  
[OI, $\lambda$6300]/H$\alpha$ and [SII, $\lambda\lambda$6716,6731]/H$\alpha$) 
for most of the galaxies studied here and plotted them on the diagnostic
diagrams of Veilleux \& Osterbrock (1987).  They found that the `eyeball' 
classifications and line--ratio based classifications agreed more than 
95\% of the time, and hence that `eyeball' classifications can be used 
with confidence to analyse large samples of 2dF spectra. 

Most of the 2dF spectra are of impressively good quality.  However, of 
the 98 spectra we examined (one galaxy was not actually observed by the 
2dFGRS), nine had such a low signal--to--noise ratio that we were unable to
classify the spectrum.  One other object appeared to be a Galactic star.  We
were therefore left with 88 good--quality 2dF spectra of candidate radio
matches with an offset D$<$10\,arcsec.  Of these 88 galaxies, 36 (41\%) were
classified as SF and 52 (59\%) as AGN.  One galaxy classified as SF had an
emission--line spectrum which resembled an AGN, but was also detected as an
IRAS source at 60\,$\mu$m.  This may be a genuinely composite object.  
Table~\ref{tab:list} lists the matched galaxies, their spectral classification, 1.4\,GHz
radio continuum flux density, apparent magnitude and redshift. A more
quantitative spectral classification using diagnostic emission--line ratios
will be presented in the forthcoming paper by Jackson and Londish (1999).

\begin{figure}[H]
\centering
\includegraphics[width=0.55\textwidth]{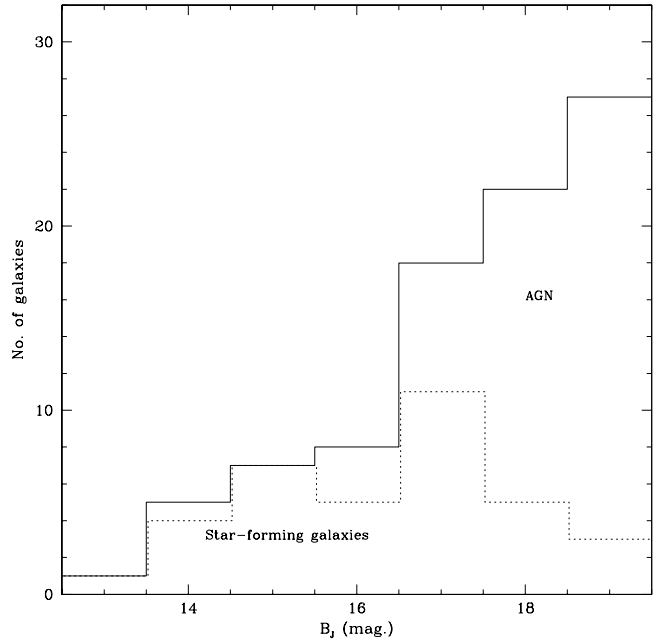}
\caption{Distribution of the star--forming and AGN galaxies 
in apparent magnitude.  Note that star--forming galaxies dominate at the 
bright end ($b_{\rm J} < 17$ mag.), while most of the fainter galaxies have 
AGN spectra.}
\label{fig:magdist}
\end{figure}
\begin{figure}[H]
\centering
\includegraphics[width=0.55\textwidth]{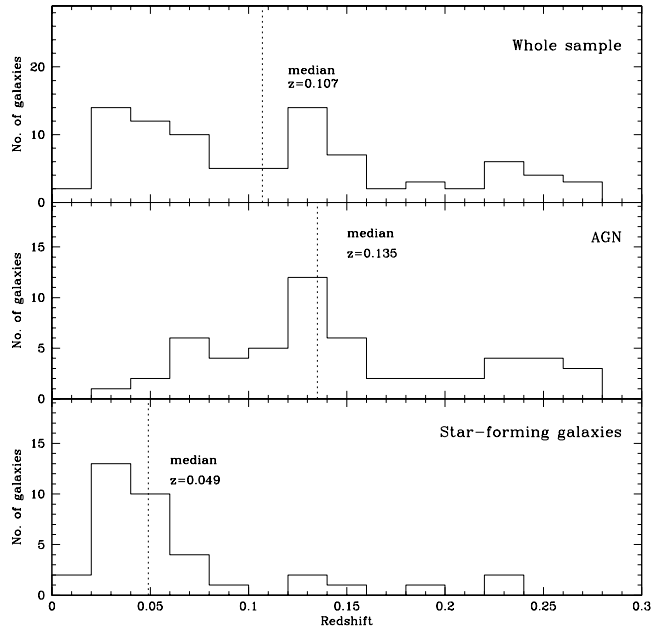}
\caption{The redshift distribution of the AGN and star--forming
galaxies, and of the whole sample. Most of the star--forming galaxies are
relatively nearby, though some extend out to redshifts of 0.25.  The AGN
galaxies are a more distant population, with a median redshift almost three
times that of the star--forming galaxies.}
\label{fig:zdist}
\end{figure}

\begin{figure}[H]
\centering
\includegraphics[width=0.6\textwidth]{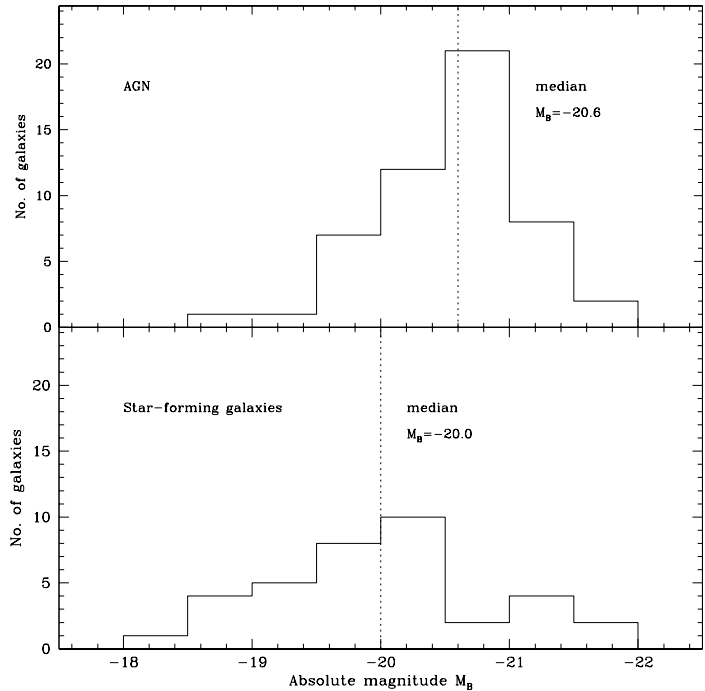}
\caption{ Absolute magnitude histograms for the AGN and 
star--forming galaxies.  The AGN spectra are typically found in luminous 
optical galaxies, while the star--forming galaxies span a much wider range 
in optical luminosity.}
\label{fig:maghisto}
\end{figure}

\begin{figure}[H]
\centering
\includegraphics[width=0.5\textwidth]{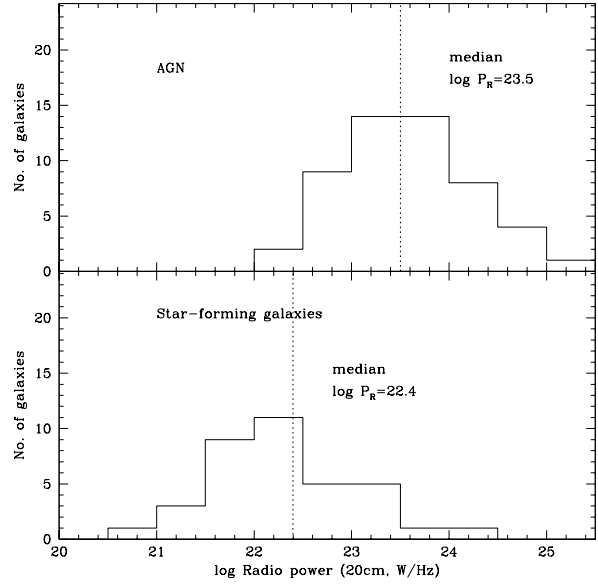}
\caption{Histograms of radio power for the AGN and star--forming galaxies.}
\label{fig:powerhisto}
\end{figure}

\begin{figure}[H]
\centering
\includegraphics[width=0.5\textwidth]{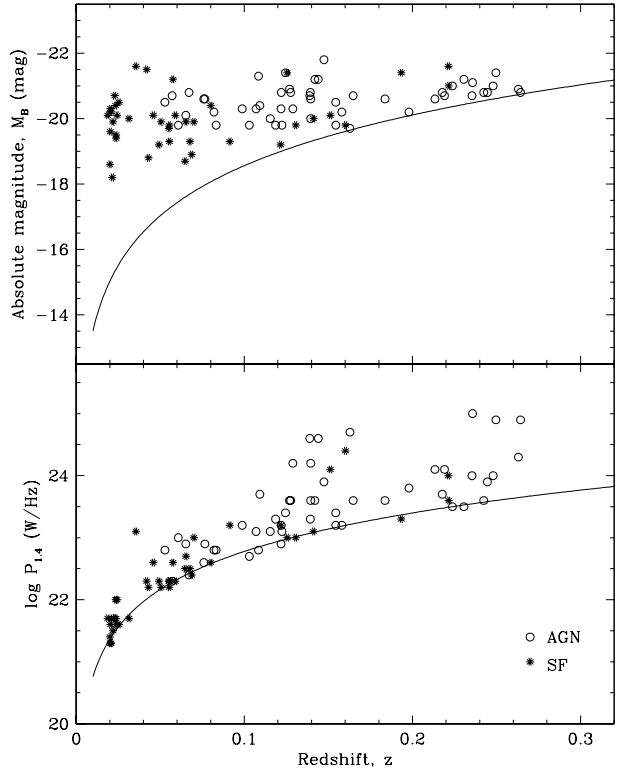}
\caption{Plots of radio power (bottom) and optical luminosity
(top) versus redshift for AGN (shown by open circles) and SF (shown by stars)
galaxies. The solid line in each plot shows the radio and optical limits 
of the NVSS and 2dFGRS respectively --- galaxies below these lines will 
drop out of our sample.}
\label{fig:powerplot}
\end{figure}

Figures~\ref{fig:magdist} and \ref{fig:zdist} show the distribution of AGN and SF classes in
apparent magnitude and redshift respectively.  There is a clear
segregation in apparent magnitude: most galaxies brighter than about
$b_{\rm J}$ = 16.5--17 magnitude fall into the star--forming (SF) class,
while the AGN class dominates the population fainter than $b_{\rm J}\sim17$.  
This reflects strong differences in the global properties of the two classes 
as well as the radio and optical flux limits of the NVSS and 2dFGRS.  
The AGN galaxies are typically more distant than the SF galaxies (by about
a factor of 3: Figure~\ref{fig:zdist}), and more luminous both optically and in radio power
(see Figures~\ref{fig:maghisto} and \ref{fig:powerhisto}). We know that the SF galaxies continue to large 
redshifts and to very faint optical magnitudes (e.g. Benn et al.\ 1993), 
but these galaxies quickly drop out of our sample because of the 2--3\,mJy 
limit of the NVSS in radio flux density.  Similarly, we know that the AGN
galaxies extend to much higher redshifts than probed by the 2dFGRS, but these
distant AGN galaxies will be fainter than the $b_{\rm J} = 19.4$ mag optical 
limit of the 2dFGRS.  Figure~\ref{fig:powerplot} shows plots of radio power and optical 
luminosity versus redshift for the AGN and SF classes --- the solid lines
correspond to the survey completeness limits of 3.5\,mJy and 19.4 mag for
the NVSS and 2dFGRS respectively.  Galaxies below these lines will be excluded
from our sample. Note that most of the SF galaxies are weak radio sources,
lying close to the NVSS cutoff at all redshifts, while most of the AGN galaxies
lie well above the radio limit but start to drop below the optical cutoff at
redshifts above 0.15. 


\begin{figure}[H]
\centering
\includegraphics[width=0.6\textwidth]{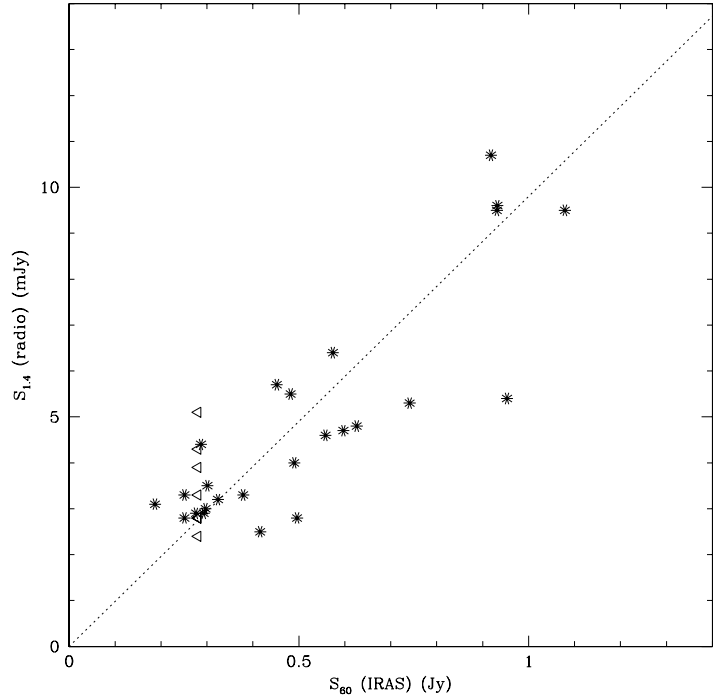}
\caption{Comparison of radio (1.4\,GHz) and IRAS far--infrared
(60\,$\mu$m) flux density for the radio--detected star--forming galaxies in the
2dFGRS. The dotted line is for $S_{60\mu{\rm m}} = 100\, S_{\rm 1.4\,GHz}$. 
Of the nine SF galaxies in Table~\ref{tab:list} which lack IRAS data, two lie in regions 
not observed by IRAS. For the remaining seven we show (open triangles) 
upper limits of 0.28\,Jy at 60$\mu$m, corresponding to the limits of the 
IRAS Faint Source Catalogue.}
\label{fig:radioir}
\end{figure}

\section{Matches with IRAS sources} 
We expect many of the SF radio sources to be IRAS detections, based on the
well--known correlation between radio continuum and far-infrared (FIR)
luminosities (e.g. Wunderlich et al.\ 1987, Condon et al.\ 1991). For spiral
galaxies, $S_{60\mu{\rm m}} \sim 100\, S_{\rm 1.4\,GHz}$ (e.g. Condon \&
Broderick 1988, Rowan--Robinson et al.\ 1993), so NVSS should detect most
galaxies in the IRAS Faint Source Catalog (which has a flux density limit of
0.28\,Jy at $60\mu{\rm m}$). 

Of the 36 galaxies classified as SF in Table~\ref{tab:list}, two (TGN222Z132 and XGN221Z023)
lie in the 3\% of the sky which has no IRAS coverage (Beichman et al.\ 1985). 
Of the remaining 34 galaxies, 27 (i.e.\ 79\%) are detected 
at 60$\mu$m in the IRAS Point Source Catalog or Faint Source Catalog (FSC). 
Figure~\ref{fig:radioir} compares the radio continuum (1.4\,GHz) and IRAS ($60\mu{\rm m}$) 
flux densities for these galaxies (for galaxies undetected by IRAS we show 
an upper limit of 0.28\,Jy, corresponding to the completeness level of the 
FSC). If we exclude one galaxy with anomalously strong 60$\mu$m emission as
discussed below, the mean FIR--radio ratio Q$_{60}$ = 
S$_{60\mu m}$/S$_{1.4\,GHz}$ for the IRAS--detected galaxies is 112$\pm$8, i.e.
close to that derived from other studies. 

One galaxy (TMS206Z015) has an unusually high value of Q$_{60}$=380, with much
stronger FIR emission than would be expected from the radio continuum flux
density.  The most likely explanation is confusion in the IRAS beam,
since this galaxy lies in a group and appears to be interacting with a
companion. 

\section{Conclusions }
Based on a study of 30 fields from the 2dF Galaxy Redshift Survey, we 
find that about 1.5\% of 2dFGRS galaxies brighter than $b_{\rm J} < 19.4$ 
magnitude are candidate identifications for 1.4\,GHz NVSS radio sources.  
Of these about 80--85\% will turn out, after closer examination, to be `real' 
associations.  Thus if the whole 2dFGRS contains 250,000 galaxies, we 
expect to identify about 4,000 candidate matches with NVSS by the time 
the 2dFGRS is complete.  About 60\% of these galaxies will be AGN 
(radio galaxies and some Seyferts) and 40\% star--forming galaxies.  
For galaxies south of declination $-30^\circ$, we will also have 843\,MHz 
radio flux density measurements from the Sydney University Molonglo Sky Survey 
(SUMSS; Bock et al.\ 1999), and will be able to measure radio spectral indices.

The final sample will be by far the largest (and most homogeneous) sample of 
radio--galaxy spectra ever obtained, and will allow us to study both the AGN 
and starburst radio populations out to redshift $z\sim0.3$, and to look for 
evidence of evolution over this redshift range.  As the present paper goes to 
press, the 2dF Galaxy Redshift Survey is 20\% complete and our sample has  
already grown to more than 700 galaxies.  This larger sample will be analysed 
in more detail in a forthcoming paper. 

\section*{Acknowledgements}

% Place acknowledgements here. Omit above \section command if there
% are no acknowledgements.

We thank the 2dF Galaxy Redshift Survey team for allowing us early access 
to their data.  We also acknowledge the essential contribution of the
many people, at the AAO and elsewhere, who have contributed to the
building and operation of the 2dF facility.  Finally, we thank the two 
anonymous referees of this paper for their careful reading and helpful 
suggestions. 

The IRAS flux densities quoted in Table~\ref{tab:list} were obtained from the NASA 
Extragalactic Database (NED).  

\section*{References}

% PASA uses the same conventions as ApJ for journal abbreviations.  Sample
% references are as follows. 
% Please follow the same format for your references.

%\reference Author, A.B. 1990 PASA 7, 2, 350

% for a journal article, or

% \reference Author, A.B. 1990 in This Is A Book Title, ed. Editor, C.D.,
% This Is A Publishers Name, 437

% for a book.
\reference 
Becker, R.H., White, R.L., Helfand, D.J. 1995 ApJ 450, 599 
\reference 
Beichman, C.A., Neugebauer, G., Habing, H.J., Clegg, P.E., Chester, T.J. 1985 
IRAS Explanatory Supplement,  JPL 
\reference 
Benn, C.R., Rowan--Robinson, M., McMahon, R.G., Broadhurst, T.J., Lawrence, A. 
1993 MNRAS 263, 98 
\reference 
Bock, D.C-J., Large, M.I., Sadler, E.M. 1999 AJ 117, 1578  
\reference 
Colless, M. 1999 Phil Trans R Soc Lond. A, 357, 105 (astro-ph/9804078) 
\reference 
Condon, J.J. 1984 ApJ 287, 461 
\reference 
Condon, J.J., Broderick, J.J. 1988 AJ 96, 30 
\reference 
Condon, J.J. 1989 ApJ 338, 13 
\reference 
Condon, J.J., Anderson, M.L., Helou, G. 1991 ApJ 376, 95 
\reference 
Condon, J.J. 1992 ARAA 30, 575 
\reference 
Condon, J.J., Cotton, W.D., Greisen, E.W., Yin, Q.F., Perley, R.A.,
Taylor, G.B., Broderick, J.J. 1998 AJ 115, 1693 
\reference 
Cress, C.M., Helfand, D.J., Becker, R.H., Gregg, M.D., White, R.L. 1996 
ApJ 473, 7  
\reference 
Folkes, S. and 24 others 1999 MNRAS, in press (astro-ph/9903456) 
\reference 
Jackson, C.A., Londish, D.M. 1999 PASA, submitted
\reference 
Lewis, I.J., Glazebrook, K., Taylor, K. 1998 SPIE Proc., 3355, 828
\reference 
Maddox, S.J. 1998 in ``Large Scale Structure: Tracks and 
Traces'', proc. 12th Potsdam Cosmology Workshop, eds. M\"{u}ller, V., 
Gottl\"{o}ber, S., M\"{u}cket, J.P., Wambsganss, J., World 
Scientific, 91 (astro-ph/9711015) 
\reference 
Magliocchetti, M., Maddox, S., Lahav, O., Wall, J.V. 1998 MNRAS 300, 257 
\reference 
Rengelink, R.B., Tang, Y., de Bruyn, A.G., Miley, G.K., Bremer, 
M.N., R\"ottgering, H., Bremer, M.A.R. 1997 A\&AS, 124, 259 
\reference 
Rowan--Robinson, M., Benn, C.R., Lawrence, A., McMahon, R.G., Broadhurst, T.J. 
1993 MNRAS, 263, 123 
\reference 
Smith, G., Lankshear, A. 1998 SPIE Proc., 3355, 905
\reference 
Veilleux, S., Osterbrock, D.A., 1987 ApJS, 63, 295 
\reference 
Windhorst, R.A., Miley, G.K., Owen, F.N., Kron, R.G., Koo, D.C. 1985 ApJ, 289,
494 
\reference 
Wunderlich, E., Klein, U., Wielebinski, R. 1987 A\&AS, 69, 487

\newpage
\begin{table}[H]

\begin{center}
\begin{tabular}{cccccrcl} 
\hline \\
  Field &  ID     & Offset  &  B(J) &  z  &S$_{1.4}$ &IRAS &  Spectral \\ 
        &         & (arcsec)& (mag) &     &(mJy) &S$_{60}$ (Jy) & class \\
\hline \\
% &&&&&&& \\
 166s& TGS166Z108 &  6.7 &  17.2 & 0.0801 &  3.0 & 0.297 & SF \\
 203s& TGS203Z019 &  2.4 &  18.9 & 0.2358 & 85.1 & & AGN (abs) \\
 203s& XGS202Z238 &  0.9 &  18.1 & 0.1389 & 90.5 & & AGN (abs) \\
 204s& TGS204Z102 &  1.8 &  16.8 & 0.0759 &  3.2 & & AGN (abs) \\
 204s& TGS204Z037 &  1.8 &  19.4 & 0.2643 & 54.0 & & AGN (E+[OIII]em) \\
 206s& TGS206Z164 &  8.8 &  14.6 & 0.0255 &  2.9 & 0.293 & SF \\
 206s& TMS206Z015 &  1.2 &  17.4 & 0.0700 & 11.3 & 4.294 & SF \\ 
 206s& XGS284Z002 &  3.7 &  17.7 & 0.0988 &  7.6 & & AGN (abs) \\
 207s& XGS206Z038 &  0.9 &  19.2 & 0.2424 &203.4 & & AGN (abs) \\
 207s& XGS208Z145 &  5.2 &  14.1 & 0.0229 &  4.7 & 0.597 & SF \\
 207s& TGS207Z011 &  0.2 &  14.2 & 0.0355 & 55.0 & & AGN (em) \\
 218s& XGS217Z158 &  3.4 &  14.4 & 0.0207 &  2.5 & 0.416 & SF \\
 218s& TGS218Z173 &  3.4 &  18.6 & 0.0914 &  8.9 & & AGN (em) \\
 218s& XGS219Z089 &  4.5 &  18.7 & 0.2497 & 59.7 & & AGN (abs) \\
 219s& TGS219Z142 &  7.3 &  17.1 & 0.1245 &  8.5 & & AGN (abs) \\ 
 220s& XGS221Z290 &  5.1 &  19.1 & 0.2480 &  7.7 & & AGN (E+[OIII]em) \\
 220s& TGS220Z015 &  0.8 &  19.3 & 0.2630 & 12.5 & & AGN (E+[OIII]em) \\
 220s& TGS220Z009 &  7.9 &  18.7 & 0.2306 &  3.1 & & AGN (E+[OII]em) \\  
 220s& TGS220Z022 &  2.8 &  15.6 & 0.0575 &  6.4 & 0.574 & SF \\
 220s& TGS220Z097 &  3.0 &  14.3 & 0.0204 &  5.4 & 0.953 & SF \\
 220s& TGS220Z128 &  6.7 &  18.9 & 0.1306 &  3.2 & & AGN (em) \\ 
 232s& TGS232Z060 &  7.6 &  16.8 & 0.0590 &  3.2 & 0.324 & SF \\
 232s& TGS232Z156 &  5.1 &  19.3 & 0.2446 &  6.1 & & AGN (abs) \\
 233s& XMS232Z027 &  6.3 &  13.2 & 0.0066 &  9.8 & 2.229 & SF \\ 
 233s& TGS233Z084 &  7.6 &  18.4 & 0.0648 &  3.8 & & AGN? ([OIII] em) \\
 233s& TGS233Z196 &  7.2 &  18.3 & 0.1030 &  2.2 & & AGN (abs) \\
 234s& TGS234Z066 &  1.8 &  18.9 & 0.1512 & 25.2 & & AGN (em) \\ 
 234s& TGS234Z197 &  6.1 &  17.9 & 0.1092 & 21.1 & & AGN (abs) \\
 234s& TGS234Z186 &  7.4 &  18.0 & 0.1393 &  4.6 & & AGN (abs) \\
 234s& TGS234Z027 &  2.3 &  14.5 & 0.0235 &  9.5 & 1.079 & SF \\
 235s& TGS235Z125 &  2.6 &  14.8 & 0.0243 &  9.6 & 0.932 & SF \\ 
 235s& XGS160Z339 &  1.3 &  18.6 & 0.1186 &  7.1 & & AGN (abs) \\
 236s& TGS236Z095 &  1.8 &  15.4 & 0.0237 &  4.6 & 0.558 & SF \\
 236s& TGS236Z091 &  4.7 &  17.0 & 0.0552 &  3.3 & 0.251 & SF \\
 236s& TGS236Z065 &  2.3 &  17.6 & 0.1420 &  9.8 & & AGN (abs) \\
 236s& TGS236Z194 &  4.7 &  15.5 & 0.0237 &  4.0 & 0.490 & SF \\   
 236s& TGS236Z171 &  3.7 &  18.5 & 0.1648 &  7.0 & & AGN (abs) \\
 237s& TGS237Z119 &  3.0 &  14.6 & 0.0419 &  5.7 & 0.452 & SF \\
 238s& TGS238Z241 &  1.6 &  18.2 & 0.2214 & 10.7 & 0.918 & SF? (or AGN?)  \\  
 238s& TGS238Z206 &  9.7 &  18.1 & 0.0008 & 15.6 & & Galactic star \\
% &&&&&&& \\
\hline
\end{tabular}
\caption{Galaxies with $\Delta < 10''$ and good--quality 
spectra}
\label{tab:list}
\end{center}
\end{table}

\newpage
\begin{table}[H]
\setcounter{table}{1} 

\begin{center}
\begin{tabular}{cccccrcl} 
\hline \\
  Field &  ID     & Offset  &  B(J) &  z  &S$_{1.4}$ &IRAS &  Spectral \\ 
        &         & (arcsec)& (mag) &     &(mJy) &S$_{60}$ (Jy) & class \\
\hline \\
% &&&&&&& \\
 238s& TGS238Z047 &  3.3 &  16.5 & 0.0215 &  5.3 & 0.741 & SF \\
 238s& TMS238Z180 &  6.9 &  17.2 & 0.1255 &  3.1 & 0.187 & SF \\
 238s& TGS238Z036 &  9.6 &  15.0 & 0.0203 &  2.8 & 0.251 & SF \\
 239s& XGS238Z030 &  3.1 &  17.2 & 0.0653 &  5.5 & 0.482 & SF \\ 
 239s& TGS239Z196 &  8.6 &  14.8 & 0.0219 &  3.5 & 0.301 & SF \\ 
 239s& TGS239Z089 &  5.4 &  18.8 & 0.2216 &  4.3 & & SF? \\
 239s& TGS239Z061 &  1.6 &  18.9 & 0.1580 &  3.3 & & AGN? \\
 240s& TGS240Z082 &  0.5 &  17.7 & 0.1269 & 11.0 & & AGN (abs+[NII]em) \\ 
 240s& TGS240Z211 &  0.6 &  18.2 & 0.1395 & 42.1 & & AGN (abs+[NII]em) \\
 240s& TGS240Z013 &  0.4 &  18.9 & 0.2238 &  3.3 & & AGN (abs+[OIII]em) \\ 
 241s& TGS241Z101 &  1.8 &  17.8 & 0.1276 & 12.7 & & AGN (abs) \\
 313s& TGS313Z081 &  4.2 &  17.7 & 0.1220 &105.8 & & AGN (abs+H$\alpha$ em) \\ 
 313s& TGS313Z100 &  7.9 &  17.1 & 0.1474 & 17.0 & & AGN (abs+H$\alpha$ em) \\ 
 314s& TGS314Z122 &  0.9 &  17.7 & 0.1440 & 98.1 & & AGN (abs) \\  
 314s& TGS314Z110 &  3.0 &  17.4 & 0.0820 &  5.1 & & AGN (abs) \\
 314s& TGS314Z116 &  5.3 &  19.3 & 0.2355 &  8.2 & & AGN (abs) \\ 
 314s& TMS314Z092 &  3.1 &  19.2 & 0.2134 & 13.2 & & AGN (abs) \\ 
 318s& TGS318Z046 &  5.8 &  18.5 & 0.1544 &  4.9 & & AGN (abs) \\
 318s& TGS318Z156 &  6.9 &  18.3 & 0.0687 &  2.8 & 0.496 & SF  \\
 319s& TGS319Z013 &  2.1 &  18.7 & 0.1224 &  4.2 & & AGN (abs) \\
 216n& TGN216Z011 &  2.2 &  18.8 & 0.1395 &  9.8 & & AGN (abs) \\
 216n& TGN216Z148 &  6.1 &  17.9 & 0.0832 &  4.1 & & AGN (abs) \\
 218n& TGN218Z083 &  5.4 &  18.1 & 0.1934 &  2.8 & & SF \\ 
 218n& TGN218Z230 &  3.1 &  15.9 & 0.0201 &  3.3 & 0.379 & SF \\ 
 218n& TGN218Z228 &  2.8 &  17.0 & 0.0652 & 10.6 & & AGN (abs) \\ 
 218n& XGN219Z136 &  1.8 &  19.4 & 0.1980 &  7.0 & & AGN (abs) \\
 220n& TGN220Z065 &  8.3 &  17.3 & 0.0492 &  4.4 & 0.287 & SF \\
 220n& TGN220Z258 &  1.4 &  17.2 & 0.0607 & 14.8 & & AGN (abs) \\
 222n& TGN222Z132 &  3.8 &  16.7 & 0.0505 &  3.5 & & SF? \\
 222n& XGN221Z023 &  2.9 &  14.3 & 0.0188 &  7.8 & & SF? \\
 222n& TGN222Z108 &  0.8 &  16.1 & 0.0571 &  3.3 & & AGN (abs+[NII]em) \\
 222n& TGN222Z318 &  3.6 &  16.1 & 0.0528 & 12.2 & & AGN (abs) \\
 222n& TGN222Z218 &  0.6 &  19.0 & 0.2178 &  5.1 & & AGN (em) \\ 
 231n& TGN231Z143 &  8.7 &  15.5 & 0.0314 &  2.9 & 0.276 & SF \\ 
 231n& TGN231Z068 &  1.6 &  18.8 & 0.1837 &  5.6 & & AGN (em) \\
 231n& TGN231Z211 &  5.5 &  16.4 & 0.0671 &  2.8 & & AGN (em) \\
 239n& TGN239Z017 &  8.0 &  18.4 & 0.1154 &  5.1 & & AGN (abs+H$\alpha$) \\
 239n& TGN239Z013 &  0.6 &  17.9 & 0.1070 &105.0 & & AGN (abs+H$\alpha$) \\
 239n& TGN239Z061 &  5.3 &  17.5 & 0.0554 &  3.3 & & SF? \\ 
 239n& TGN239Z082 &  9.6 &  18.8 & 0.1412 &  2.8 & & SF \\
% &&&&&&& \\
\hline
\end{tabular}
\caption{Galaxies with $\Delta < 10''$ and good--quality 
spectra (contd.) }
\end{center}
\end{table}

\newpage
\begin{table}[H]
\setcounter{table}{1} 

\begin{center}
\begin{tabular}{cccccrcl} 
\hline \\
  Field &  ID     & Offset  &  B(J) &  z  &S$_{1.4}$ &IRAS &  Spectral \\ 
        &         & (arcsec)& (mag) &     &(mJy) &S$_{60}$ (Jy) & class \\
\hline \\
% &&&&&&& \\
 239n& XGN238Z202 &  0.9 &  19.3 & 0.1602 & 46.3 & & AGN? (em) \\
 239n& TGN239Z221 &  2.8 &  17.4 & 0.0430 &  4.8 & 0.626 & SF \\ 
 239n& TGN239Z171 &  1.4 &  18.2 & 0.1219 &  2.6 & & AGN (abs+[NII]) \\
 239n& TGN239Z158 &  9.8 &  18.3 & 0.1289 & 52.5 & & AGN (abs) \\
 239n& TGN239Z172 &  3.7 &  16.2 & 0.0459 &  9.5 & 0.931 & SF, interacting? \\ 
 240n& TGN240Z127 &  8.1 &  17.0 & 0.0554 &  2.4 & & SF?? \\
 240n& XGN309Z233 &  4.6 &  17.9 & 0.0677 &  3.9 & & SF? \\ 
 240n& TGN240Z277 &  0.6 &  16.9 & 0.0766 &  6.4 & & AGN (abs) \\
 240n& TGN240Z219 &  8.5 &  19.3 & 0.1216 &  5.1 & & SF? \\
% &&&&&&& \\
\hline
\end{tabular}
\caption{Galaxies with $\Delta < 10''$ and good--quality 
spectra (contd.) }
\end{center}
\end{table}

\end{document}